\documentclass[10pt, twocolumn, nofootinbib,superscriptaddress]{revtex4-1}
\usepackage{amsmath,amssymb,amsfonts}
\usepackage{algorithmic}
\usepackage{graphicx}
\usepackage{textcomp}
\usepackage{xcolor}
\usepackage{ragged2e}
\usepackage{booktabs, makecell, tabularx}

\usepackage{gensymb}
\makeatletter
    \renewcommand\@make@capt@title[2]{%
     \@ifx@empty\float@link{\@firstofone}{\expandafter\href\expandafter{\float@link}}%
      {\textbf{#1}}\@caption@fignum@sep#2\quad}%
\makeatother
\makeatletter 
\renewcommand{\fnum@figure}{\textbf{Fig.~\thefigure}}
\makeatother
\usepackage{xcolor}

\def\BibTeX{{\rm B\kern-.05em{\sc i\kern-.025em b}\kern-.08em
    T\kern-.1667em\lower.7ex\hbox{E}\kern-.125emX}}
\begin{document}

\title{Observation of a Brillouin dynamic grating in silicon nitride waveguides}

\author{Roel~Botter$^\dag$}
\author{Jasper~van~den~Hoogen$^\dag$}
\author{Akhileshwar~Mishra}
\author{Kaixuan~Ye}
\affiliation{Nonlinear Nanophotonics, MESA+ Institute of Nanotechnology, University of Twente, Enschede, the Netherlands}
\author{Albert~van~Rees}
\affiliation{Laser Physics and Nonlinear Optics, MESA+ Institute of Nanotechnology, University of Twente, Enschede, the Netherlands}
\author{Marcel~Hoekman}
\affiliation{LioniX International, Enschede, the Netherlands}
\author{Klaus~Boller}
\affiliation{Laser Physics and Nonlinear Optics, MESA+ Institute of Nanotechnology, University of Twente, Enschede, the Netherlands}
\author{David~Marpaung}
\email{Corresponding author: david.marpaung@utwente.nl}
\affiliation{Nonlinear Nanophotonics, MESA+ Institute of Nanotechnology, University of Twente, Enschede, the Netherlands}

\begin{abstract}
    Brillouin enhanced four wave mixing in the form of a Brillouin dynamic grating (BDG) enables a uniquely tunable filter, whose properties can be tuned by purely optical means. This makes the BDG a valuable tool in microwave photonics (MWP). BDGs have been studied extensively in fibers, but the only observation in an integrated platform required exotic materials. Unlocking BDG in a standard and mature platform will enable its integration into large-scale circuits. Here we demonstrate the first observation of a BDG in a silicon nitride (Si$_3$N$_4$) waveguide. We also present a new, optimized design, which will enhance the BDG response of the waveguide, unlocking a path to large-scale integration into MWP circuits.
\end{abstract}

\maketitle
\def\thefootnote{\dag}\footnotetext{These authors contributed equally to this work}

\section*{Introduction}

Optomechanical interactions have been gaining interest as a research topic recently \cite{Aspelmeyer_RevModPhys_2014, Safavi_Opt_2019, Eggleton_NatPhot_2019}, showing promising applications in many fields such as telecommunication \cite{Li_NatComm_2013, Marpaung_Optica_2015}, quantum technologies \cite{Barzanjeh_NatPhys_2022}, high resolution spectrometers \cite{Dong_OL_2014} and sensing \cite{Kim_JLT_2015, Li_Opt_2017, Li_Optica_2018}.

A Brillouin dynamic grating (BDG) introduces enhanced flexibility into optomechanics. This grating is unique because it can be tuned in both strength and bandwidth via purely optical means. Furthermore, the probe light is orthogonally polarized from the pumps, so a polarization beam splitter can be used to combine and split the signals. BDG has shown promising applications in microwave photonics (MWP) such as variable time delays \cite{Chin_LPR_2012, Merklein_JoO_2018}, multi-tap filters \cite{Sancho_OE_2012} and true-time reversal \cite{Santagiustina_SciRep_2013}.

BDGs have first been demonstrated in polarization maintaining fiber \cite{Song_OL_2008}, and later also in single-mode \cite{Song_OL_2011} and few-mode fibers \cite{Li_OL_2012}. However the only on-chip observation so far has been in a chalcogenide, As$_2$S$_3$, based waveguide \cite{Pant_OL_2013}. Although this material is a great host for Brillouin scattering, its integrability is limited, leading to a reduced potential for use in large-scale circuits.

In this work, we show for the first time the observation of a BDG in a standard, low-loss silicon nitride (Si$_3$N$_4$) waveguide. We build on our recent work where we showed guided-acoustic Brillouin scattering in the multilayer silicon nitride platform, leading to record-high Brillouin gain coefficients in Si$_3$N$_4$ waveguides \cite{Botter_SciAdv_2022}. This is a substantial advancement in the application of Brillouin dynamic gratings in microwave photoncs, as it unlocks the process in a mature photonics platform that has a large library of MWP components, in which we have recently shown record-high performance \cite{Roeloffzen_JSTQE_2018, Daulay_NatComm_2022, Liu_APLPhot_2023}.

\section*{Operating principle}

\begin{figure}
    \centering
    \includegraphics[width=\linewidth]{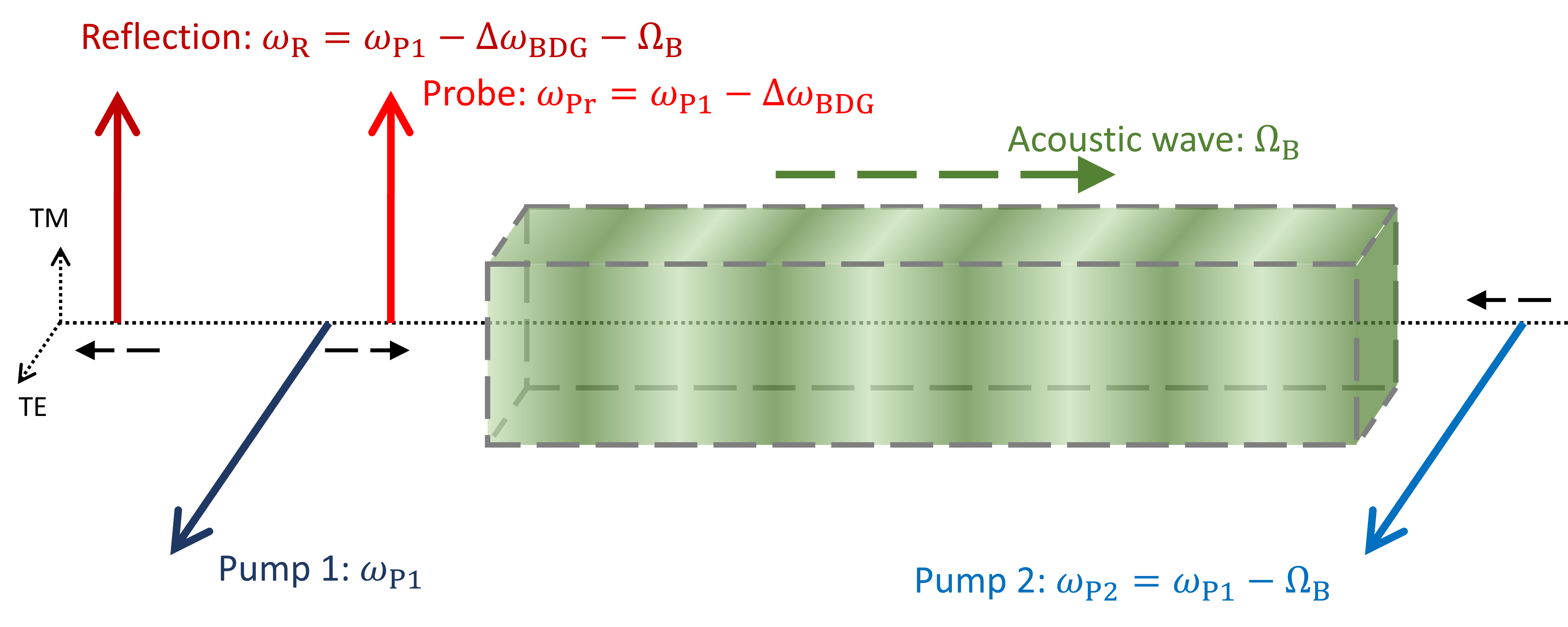}
    \caption{\textbf{Schematic depiction of the principle of a Brillouin dynamic grating.} The interaction between pump 1 and pump 2 (in blue) in the TE polarization creates an acoustic wave with frequency $\Omega_\mathrm{B}$ via SBS. The probe wave (in red), in the TM polarization, is reflected by the acoustic wave, undergoing a Doppler shift in the process. Pump 1 and probe are separated by $\Delta\omega_{\mathrm{BDG}}$.}
    \label{fig:BDG-schem}
\end{figure}

The creation and probing of Brillouin dynamic gratings is a four wave mixing process that is supported by stimulated Brillouin scattering (SBS), and therefore represents a form of Brillouin enhanced four wave mixing \cite{Bergman_Sensors_2018}. A schematic depiction of the process can be seen in Fig.~\ref{fig:BDG-schem}. In essence a BDG is an extension of the SBS process with an additional probe wave.

In SBS a probe wave interacts with a counter-propagating higher power pump wave, which is detuned with a frequency on the order of 10~GHz. If the detuning of these two waves matches the Brillouin frequency of the waveguide, $\Omega_\mathrm{B}$, the interference between these two waves creates a sound wave through electrostriction. The frequency of this sound wave is equal to the Brillouin frequency. Via the photoelastic effect this sound wave then creates a moving Bragg-like grating. This grating will reflect the pump light, which is then Doppler shifted due to the motion of the acoustic grating. The reflected pump light will then match the frequency of the probe, resulting in a amplification with a typical bandwidth on the order of 10 -- 100~MHz \cite{Boyd_Book_2008, Agrawal_Book_2013}.

For the extension of the SBS process into the generation of a BDG, the probe wave is replaced by a second pump, as depicted in Fig.~\ref{fig:BDG-schem}. Both pumps are in the transverse electric (TE) polarization, and generate an acoustic wave as described above. An new probe wave, transverse magnetic (TM) polarized and therefore orthogonal to the two pumps, is then introduced, co-propagating with the higher-power, higher-frequency pump~1. This probe wave is then used to characterize the grating created by the acoustic wave. When the probe light has the right wavelength, it will be reflected by the grating, undergoing a Doppler shift equal to the Brillouin frequency $\Omega_\mathrm{B}$.

Light will only be reflected by the Bragg-like acoustic grating if the wavelength in the waveguide matches the grating period. In the waveguide the wavelength of pump~1 is  

\begin{equation}
    \lambda_{\mathrm{P1,wg}} = \frac{\lambda_{\mathrm{P1,0}}}{n_{\mathrm{eff,TE}}}
\end{equation}

and the wavelength of the probe is

\begin{equation}
    \lambda_{\mathrm{Pr,wg}} = \frac{\lambda_{\mathrm{Pr,0}}}{n_{\mathrm{eff,TM}}}. 
\end{equation}

Here $\lambda_{\mathrm{P1,0}}$ and $\lambda_{\mathrm{Pr,0}}$ are the pump 1 and probe vacuum wavelengths respectively, and $n_{\mathrm{eff,TE}}$ and $n_{\mathrm{eff,TM}}$ are the effective refractive indices of the TE and TM mode of the waveguide.

In the SBS process between the two pumps, pump~1 is reflected by the acoustic grating. This means that the wavelength of pump~1 is in the reflection band of the grating, and for the probe wavelength to reflect it should be the same. Using this we can calculate the required probe wavelength: 

\begin{align}
    \lambda_{\mathrm{Pr,wg}} &= \lambda_{\mathrm{P1,wg}}\\
    \frac{\lambda_{\mathrm{P1,0}}}{n_{\mathrm{eff,TE}}} &= \frac{\lambda_{\mathrm{Pr,0}}}{n_{\mathrm{eff,TM}}}\\[2.5pt]
    \lambda_{\mathrm{Pr,0}} &= \frac{n_{\mathrm{eff,TM}}\lambda_{\mathrm{P1,0}}}{n_{\mathrm{eff,TE}}}.\label{eq:wl}
\end{align}

This equation shows one of the advantages of a BDG, compared to the regular Brillouin process. In a BDG the frequency separation of pumps and probe, $\Delta\omega_{\mathrm{BDG}}$, can be up to tens of nanometers, depending on the birefringence of a waveguide, which is much larger than the 10~GHz typically seen in SBS. This makes it easier to remove the pump light from the signal path, which reduces noise that can be introduced by the high power of the pump waves. Furthermore, the difference in polarization enables an extra filtering step for separating pump and probe with the use of a polarizer or a polarization beam splitter, strengthening this effect.

The biggest advantage of a BDG is its unique dynamic nature. Assuming the grating is weak, which means the refractive index variation is small, the bandwidth $\Delta\omega$ of a BDG is determined by \cite{Dong_OE_2010}

\begin{equation}\label{eq:bw}
    \Delta\omega = \frac{\pi c}{n_{\mathrm{eff,TM} L}}
\end{equation}

where $c$ is the speed of light, and $L$ the grating length. The length of the grating can be tuned by pulsing the pump waves, as the acoustic wave is only generated where the pump waves overlap. This way the bandwidth of the grating can be wider or narrower by making the grating shorter or longer. Furthermore, the time delay between the input probe and reflection can also be tuned by changing the location where the pump pulses meet and overlap, thus the location of the grating. These two dynamically tunable parameters of the grating response make the BDG a uniquely agile filter.

\section*{Waveguide properties}

\begin{figure}
    \centering
    \includegraphics[width=\linewidth]{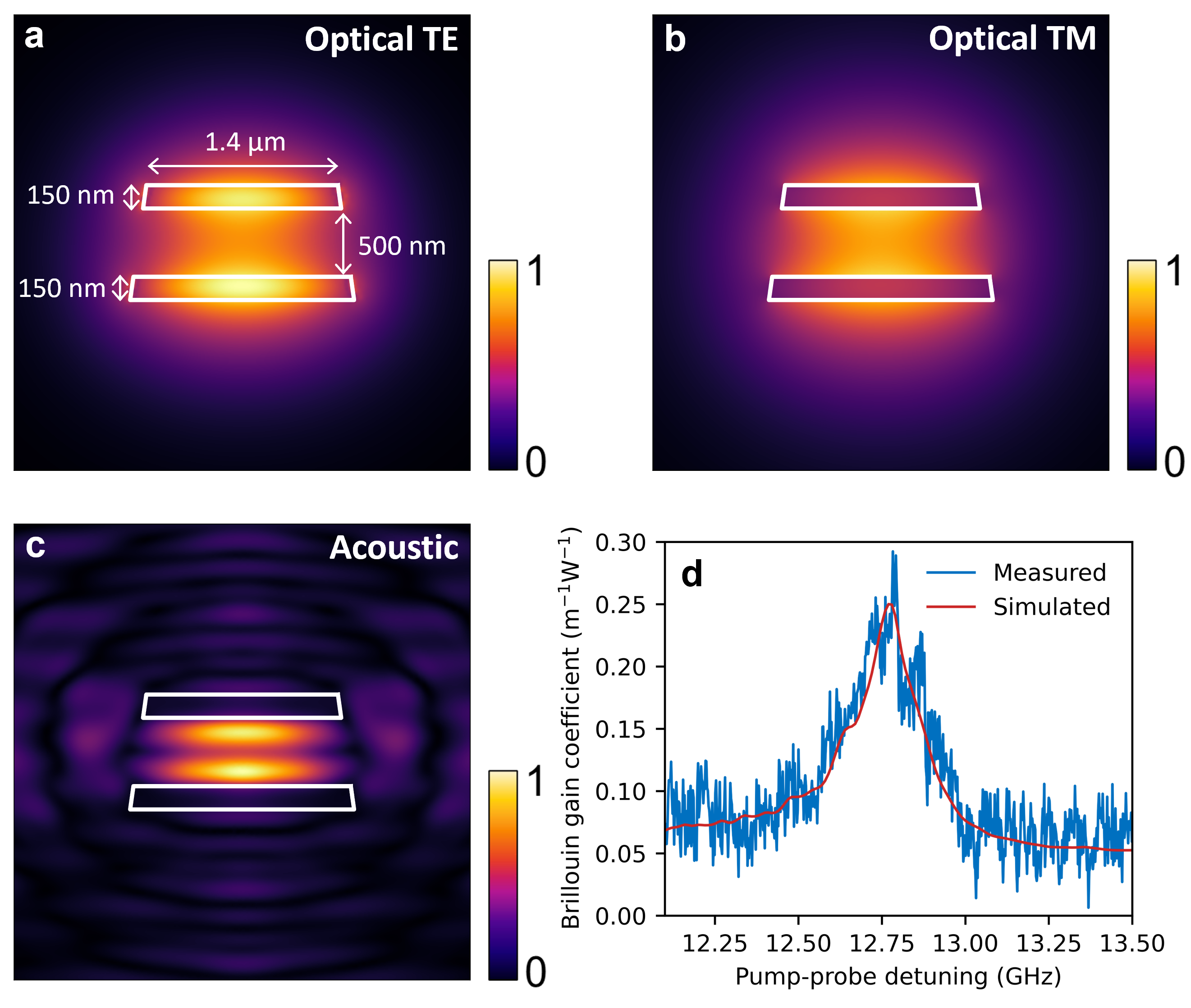}
    \caption{\textbf{The different mode fields of the waveguide.} \textbf{a} The optical TE mode at 1556~nm with the dimensions of the waveguide overlayed. \textbf{b} The optical TM mode at 1502~nm. \textbf{c} The acoustic response at 12.76~GHz. \textbf{d} The measured and simulated Brillouin gain coefficient of the waveguide at 1550~nm, from \cite{Botter_SciAdv_2022}.}
    \label{fig:mode_fields}
\end{figure}

In this research we used silicon nitride waveguides. They are made using the symmetric double stripe (SDS) geometry as shown in Fig.~\ref{fig:mode_fields}a, which consists of two silicon nitride (Si$_3$N$_4$) layers which are both 150~nm high, and separated by a 500~nm layer of silicon oxide (SiO$_2$) \cite{Botter_SciAdv_2022, Roeloffzen_JSTQE_2018}. A selection of widths is available in our samples, these are listed in Table~\ref{tab:tab1}, along with their properties. These waveguides have been optimized for single-mode use of the TE$_{00}$ mode, with low-loss chip-to-fiber coupling and the lowest bending radius that does not increase the propagation losses. The TM mode is not confined as much as the TE mode, which is reflected in the lower TM refractive indices. As a result the losses in the chip-to-fiber coupling and in the bends is significantly higher. 

\begin{table}
    \centering
    \caption{\textbf{The parameters of the available waveguides, depending on their widths $w_{\mathrm{wg}}$.} The simulated effective indices $n_{\mathrm{eff,TE}}$ of the TE mode are for a pump 1 wavelength of 1556~nm, the effective indices $n_{\mathrm{eff,TM}}$ of the TM mode is simulated for the wavelength $\lambda_{\mathrm{Pr}}$ that will match the pump 1 according to Eq~\eqref{eq:wl}. The Brillouin gain coefficients $g_{\mathrm{B}}$ and TE propagation losses $\alpha_{\mathrm{TE}}$ are measured values \cite{Botter_SciAdv_2022}.}
    \begin{tabular}{c|c|c|c|c|c}
        $w_{\mathrm{wg}}$ & \multicolumn{2}{c|}{$n_{\mathrm{eff}}$} & $\lambda_{\mathrm{Pr}}$ & $g_{\mathrm{B}}$ & $\alpha_{\mathrm{TE}}$ \\
        (µm) & [TE] & [TM] & (nm) & (m$^{-1}$W$^{-1}$) & (dB/cm) \\
        \hline
        1.1 & 1.528 & 1.482 & 1509 & 0.20 & 0.228 \\
        1.2 & 1.534 & 1.485 & 1506 & 0.24 & 0.223 \\
        1.4 & 1.543 & 1.490 & 1502 & 0.25 & 0.206 \\
        1.5 & 1.547 & 1.492 & 1501 & 0.25 & 0.230 \\
        3.0 & 1.574 & 1.509 & 1492 & 0.40 & 0.195
    \end{tabular}
    \label{tab:tab1}
\end{table}

To maximise the observable signal we want to use a waveguide that has a high Brillouin gain and low (TE) propagation losses, and the birefringence should not be too high, as our probe laser has a lower wavelength limit of about 1500~nm. Based on these criteria and the data from Table~\ref{tab:tab1} we selected the 1.4~µm wide waveguide, which has the highest Brillouin gain coefficient and propagation loss for the waveguides with a simulated probe wavelength of more than 1500~nm.

In this work we used an SDS waveguide with a length of 50~cm, folded in a quasi-rectangular spiral. The wavelength of the pump laser is around 1556~nm, which in the TE mode corresponds to an effective index of 1.543, with a mode field as shown in Fig.~\ref{fig:mode_fields}a. The corresponding probe wavelength calculated combining Eq.~\ref{eq:wl} with mode simulations in \textit{Optodesigner} is 1502~nm, with an effective TM index of 1.490, shown in Fig.~\ref{fig:mode_fields}b. In this waveguide the TE mode has a Brillouin gain of 0.25~m$^{-1}$W$^{-1}$ with a shift of 12.76~GHz, a propagation loss of 0.206~dB/cm and a coupling loss of 1.3~dB/facet \cite{Botter_SciAdv_2022}. The acoustic mode at the peak Brillouin shift is depicted in Fig.~\ref{fig:mode_fields}c. Simulations show that for the TM mode the fiber-chip coupling losses are about 6.5 dB/facet, and the tight bends (radius 100~µm) will have losses of 82~dB/cm, which corresponds to 1.2~dB per 90° bend. As a result, the TM mode will not propagate far into the waveguide, leading to a low reflection, and a weak BDG signal. 

The high TM losses do lead to a shorter effective grating length, and as a result a larger bandwidth. According to Eq.~\eqref{eq:bw} the bandwidth is 1.2~GHz if it is present in the full 50~cm length of our waveguide spiral. However, if we assume only the 1~cm of waveguide between the facet and the first bend contributes, the bandwidth is 61~GHz. This is a slight overestimation, but it shows that the observation will be easier, as the probe laser can be scanned with larger steps.

\section*{Experimental setup}

\begin{figure}
    \centering
    \includegraphics[width=\linewidth]{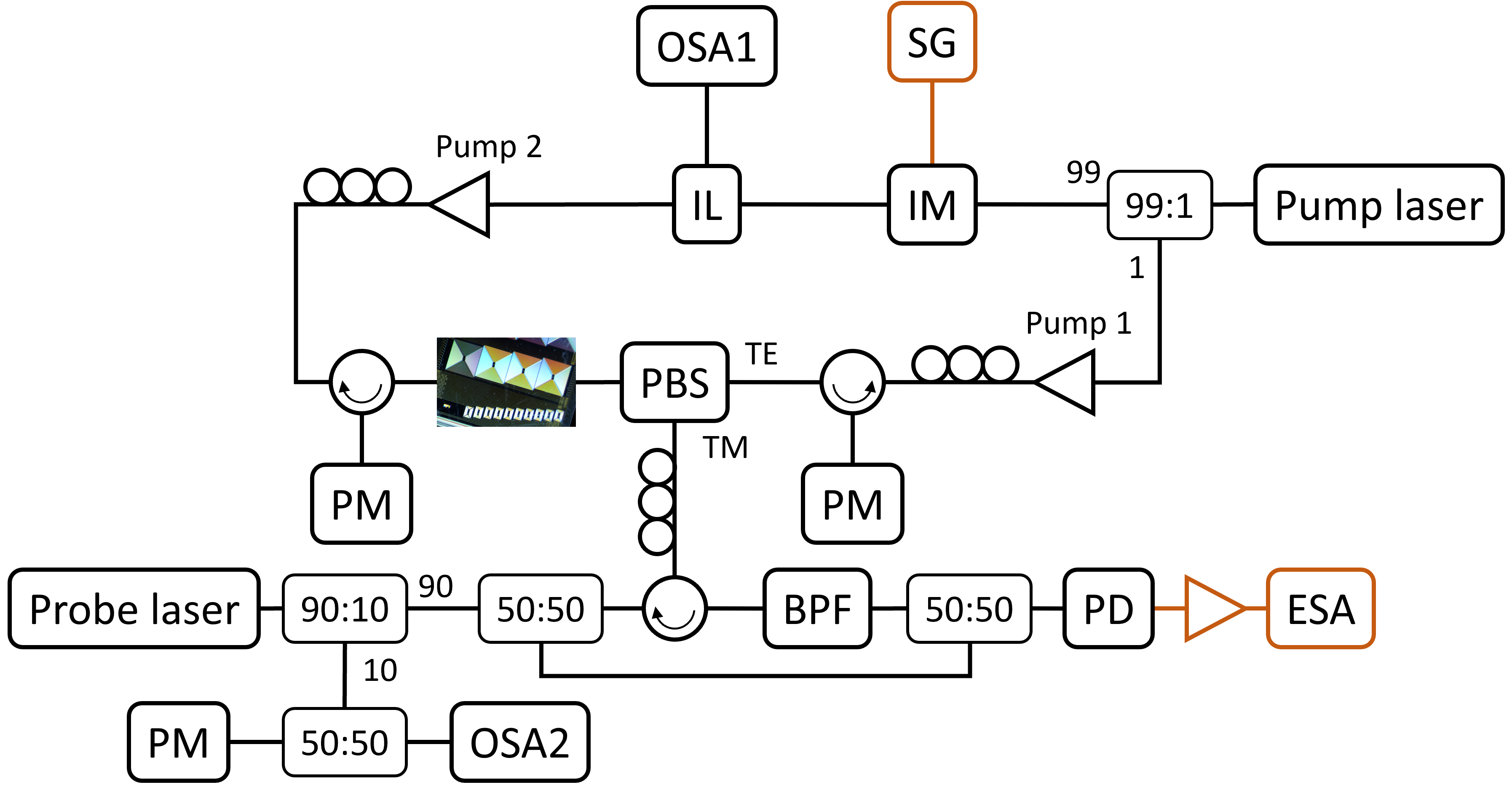}
    \caption{\textbf{The setup used in this work.} BPF: band-pass filter,  ESA: electrical spectrum analyser, IM: intensity modulator, IL: interleaver, OSA: optical spectrum analyzer, PBS: polarization beam splitter, PD: photodiode, PM: power meter, SG: signal generator.}
    \label{fig:setup}
\end{figure}

Fig.~\ref{fig:setup} shows the setup used in this experiment. For mutual frequency stability the two pump waves are derived from the same laser source, an APIC Ultra Low Noise Laser. The laser output is split using a 99:1 splitter to create the two pump lines. This laser is set to 1555.82~nm, to match the pump wavelengths with the interleaver response.

\begin{figure}
    \centering
    \includegraphics[width=\linewidth]{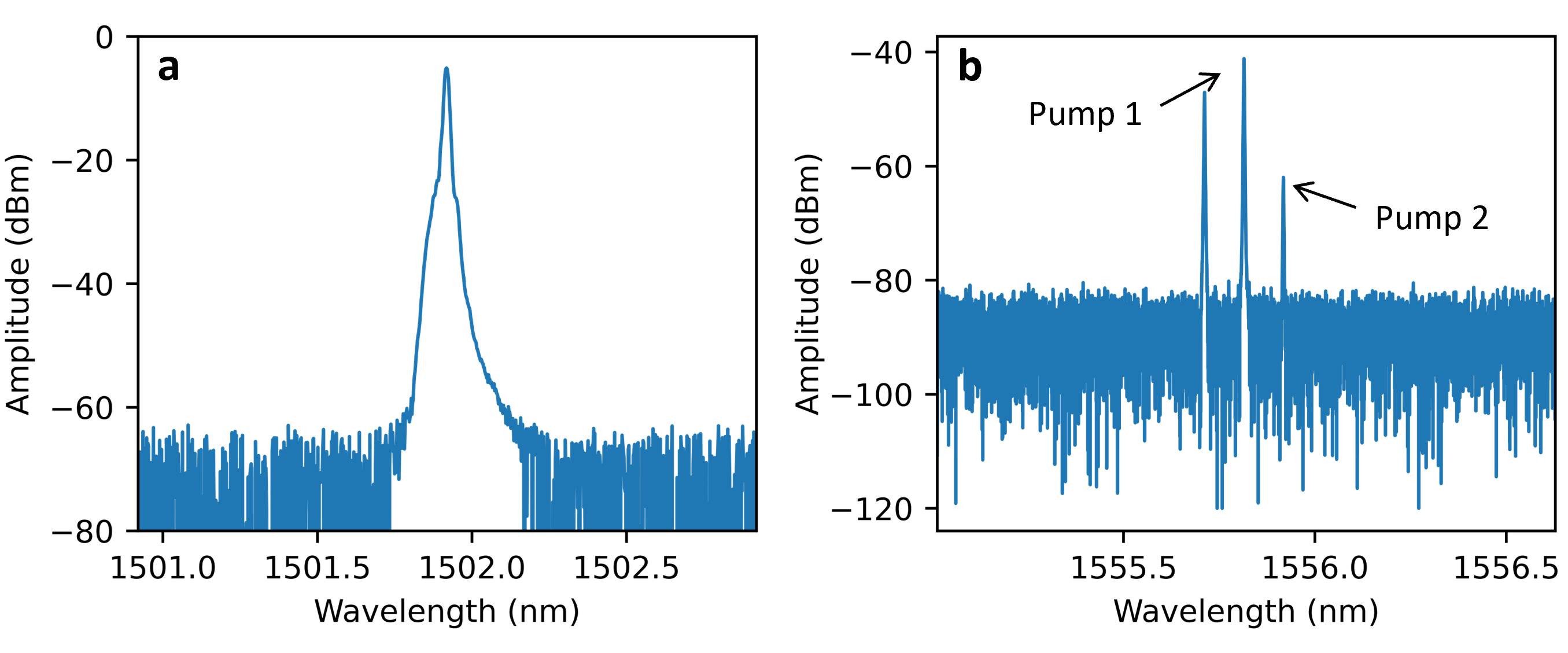}
    \caption{\textbf{The measurements of the laser wavelengths.} \textbf{a} The measurement of the pump wavelengths, on OSA1 in Fig.~\ref{fig:setup}. Pump 2 is attenuated due to the interleaver response, and pump 1 and the lower sideband created in the IM are clearly visible. \textbf{b} The measurement of the probe wavelength, on OSA 2 in Fig.~\ref{fig:setup}}
    \label{fig:lasers}
\end{figure}

The 99\% arm is modulated using a Thorlabs LNA6213 intensity modulator, connected to a Wiltron 69147A signal generator set to 12.76~GHz, the Brillouin shift frequency of the waveguide. The lower sideband is then selected using an Optoplex IL-CTBFAC687 interleaver, and amplified to 21.4~dBm using an Amonics AEDFA-PA-35-B-FA, creating pump~2. The upper sideband and carrier are sent to a Finisar 1500s optical spectrum analyser (labelled OSA1 in Fig.~\ref{fig:setup}) to determine the pump frequencies. The measured spectrum is shown in Fig.~\ref{fig:lasers}b.

The 1\% arm of the splitter is ampilified using an Amonics AEDFA-33-B-FA to 33.8~dBm, and serves as pump~1.

\begin{figure}
    \centering
    \includegraphics[width=0.8\linewidth]{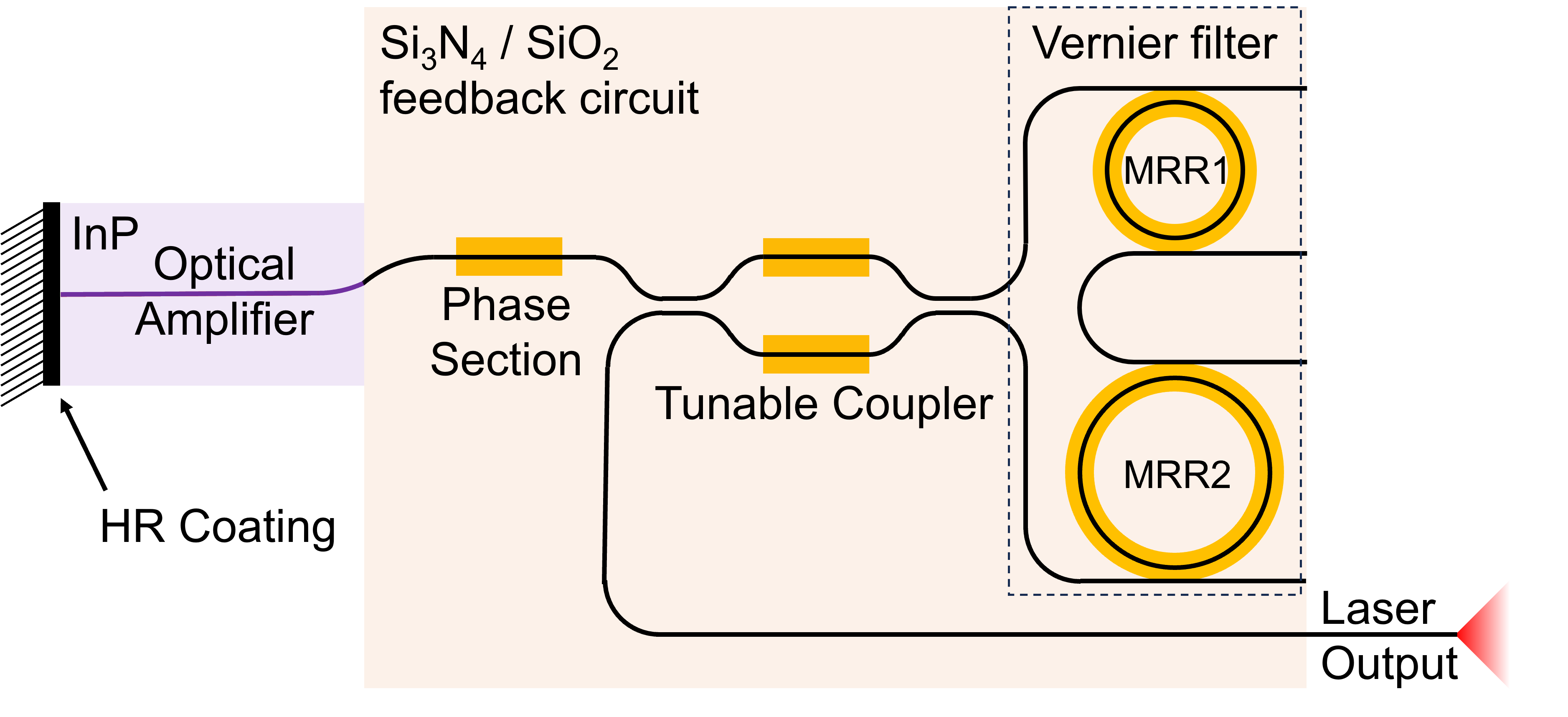}
    \caption{\textbf{Schematic depiction of the probe laser.} The indium phosphide optical amplifier is hybirid integrated with a silicon nitride feedback chip. This feedback chip provides an extended cavity as well as a tunable frequency selective mirror, in the form of a Vernier filter consisting of two microring resonators (MRR). The other mirror is formed by the highly reflective (HR) coating on the amplifier. A tunable coupler provides the output light.}
    \label{fig:laser}
\end{figure}

To reach the required probe wavelength we use a hybrid integrated laser, consisting of an indium phosphide gain section and a silicon nitride feedback circuit, as depicted in Fig.~\ref{fig:laser} \cite{Oldenbeuving_LPL_2013, Fan_IEEEPhot_2016, VanRees_OE_2020}. This Vernier filter of this laster is designed a tuning range of 120~nm around 1540~nm. This laser is tuned to the BDG frequency, outputting 8.59~dBm. 10\% of the light is split off to a power meter and an Ando AQ6317 optical spectrum analyzer to measure the wavelength, shown in Fig.~\ref{fig:lasers}a. The other 90\% of light of this laser is split 50:50, with one half acting as the probe, and the other half used for the heterodyne detection of the BDG reflection.

The probe light (in the TM mode) is combined with the light of pump 1 (in the TE mode) using a polarization beam splitter (PBS). This light is then sent into the chip. Pump 2 is sent to the waveguide from the other direction, also in the TE mode.

Part of the probe light is reflected in the BDG process. This reflected light is filtered from the pumps via the PBS and split off via a circulator. The signal is then further filtered using an EXFO XTM-50-SCL-U tunable bandpass filter, to block any pump light that has passed the PBS. The reflected light is then combined with the probe light for heterodyne detection on a Discovery Semiconductor DSC30S photodiode. This signal is amplified with a Mini-Circuits ZVA-213-S+ gain block. The RF signal is then analyzed using a Keysight N9000B electrical spectrum analyser (ESA).

\section*{Results}

\begin{figure}
    \centering
    \includegraphics[width=0.8\linewidth]{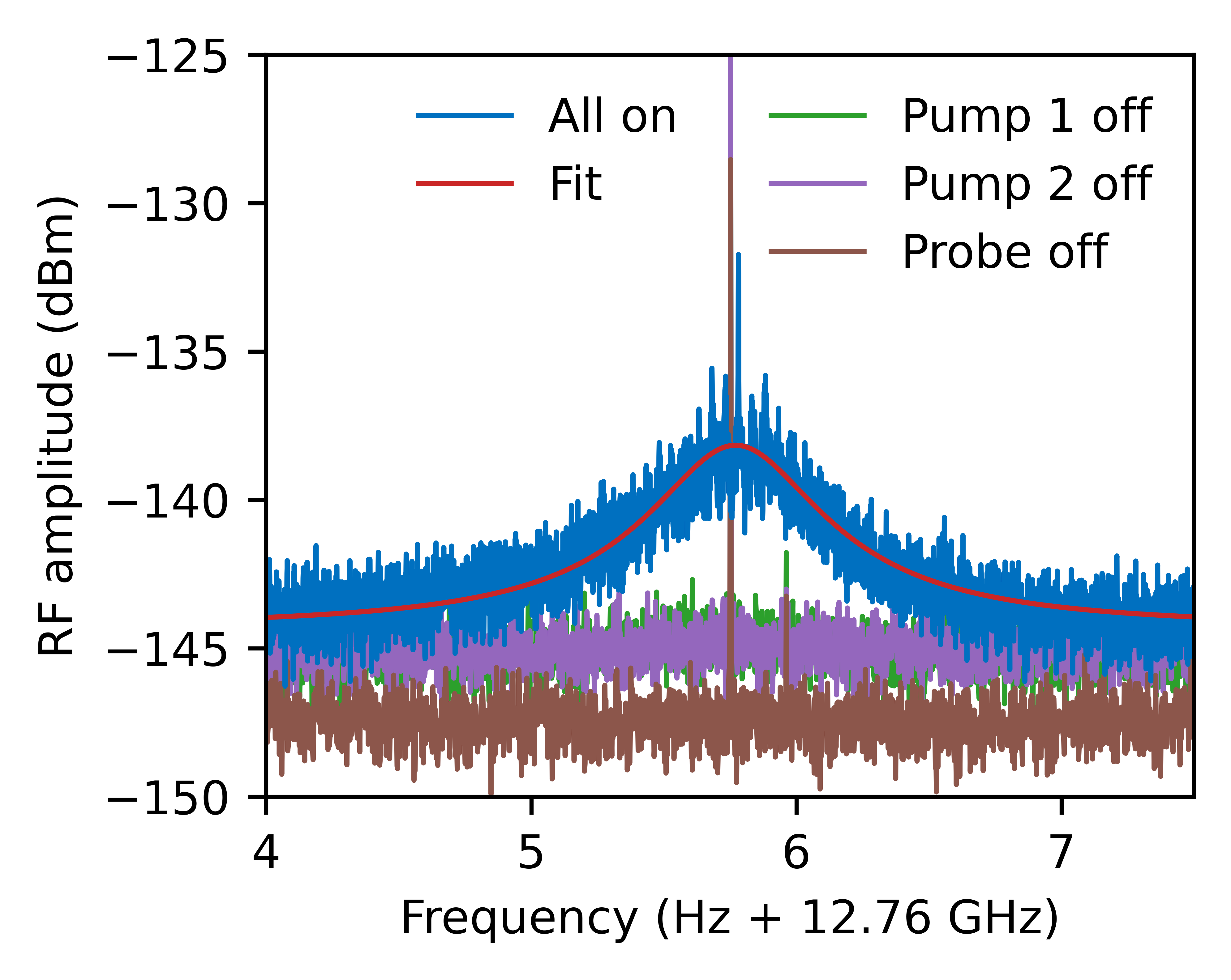}
    \caption{\textbf{The measured BDG signal, overlayed with a fit of the response.} The signal is only present when all three light sources are active, confirming that it is the result of a BDG interaction. The center of the signal trace shows a large spike due to interference from the signal generator, which is present during every measurement.}
    \label{fig:onoff}
\end{figure}

\begin{figure}
    \centering
    \includegraphics[width=0.8\linewidth]{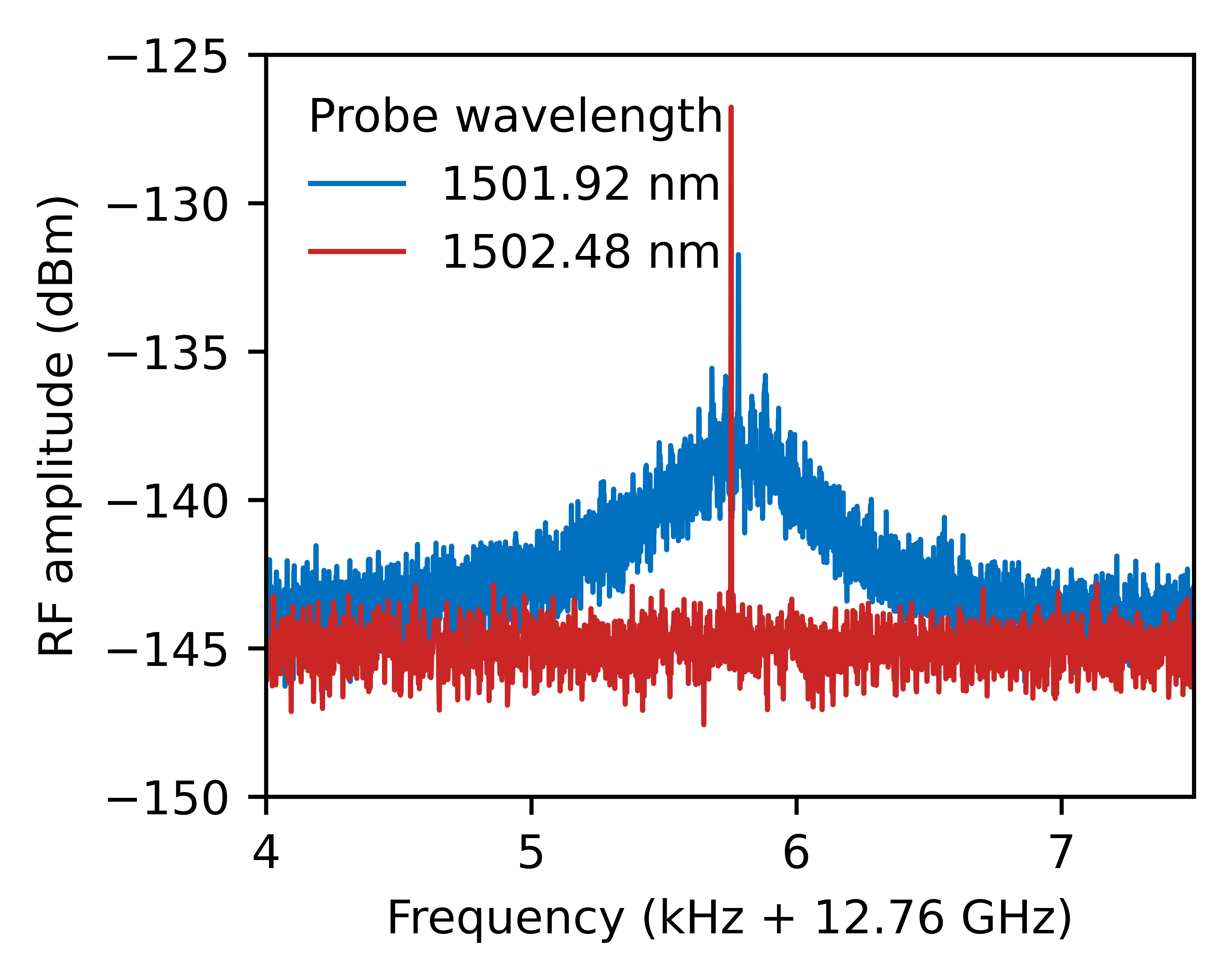}
    \caption{\textbf{The bandwidth of the BDG signal.} When the probe is moved by 0.5~nm the BDG signal is no longer observed. This is another confirmation that the signal is caused by a BDG and not a different interaction, which would have a broader bandwidth.}
    \label{fig:move}
\end{figure}

After scanning the probe laser around 1502~nm we found a response when the probe was at 1501.92~nm, as can be seen in Fig.~\ref{fig:onoff}. To verify that the peak is the result of a BDG interaction, we repeated the measurement with each time one of the three light sources turned off. In each of these cases there was no signal present, as depicted in Fig.~\ref{fig:onoff}. A second verification step was to change the probe wavelength by more than 61~GHz, our largest estimate of the BDG bandwidth. As expected, the signal was also not present when the probe wavelength was tuned to 1502.48~nm, as shown in Fig.~\ref{fig:move}. 

The peak is not exactly at 12.76~GHz due to a small frequency offset between the signal generator and the ESA. The measurement shows one large spike due to interference from the signal generator to the spectrum analyser. A Lorentzian fit using the \textit{Python} package \textit{lmfit} is overlayed, showing a full width at half maximum of 603~Hz.

From the fit in Fig.~\ref{fig:onoff} we can see that the recorded signal has an RF power of -138.1~dBm. The gain block has an amplification of 26~dB, meaning that the actual RF signal is at -164.1~dBm.

The photodiode generates a photocurrent $I_{\mathrm{PD}}(t)$ due to the incoming optical field $E_{\mathrm{PD}}(t)$ equal to

\begin{equation}
    I_{\mathrm{PD}}(t) = \frac{1}{2}\mathcal{R}_{\mathrm{PD}}E_{\mathrm{PD}}(t)E_{\mathrm{PD}}^*(t)
\end{equation}

Where $\mathcal{R}_{\mathrm{PD}}$ is the responsivity of the photodiode, in this case 0.8~A/W. The resulting RF power at the ESA, assuming matched loads, is then equal to

\begin{align}
    \begin{split}
        P_{\mathrm{RF}} &=\frac{1}{4}R_\mathrm{L}\langle I_{\mathrm{PD}}(t)^2\rangle\\
        &= \frac{1}{16} R_\mathrm{L}\mathcal{R}_{\mathrm{PD}}{}^2\langle E_{\mathrm{Pr}}(t)E_{\mathrm{R}}(t)^*\rangle \langle E_{\mathrm{R}}(t)E_{\mathrm{pr}}(t)^*\rangle\\
        &= \frac{1}{16} R_\mathrm{L}\mathcal{R}_{\mathrm{PD}}{}^2 P_{\mathrm{Pr}}P_{\mathrm{R}}
    \end{split}
\end{align}

with $R_\mathrm{L}$ the input impedance of the ESA, which is 50~$\Omega$, and $P_{\mathrm{Pr}}$ and $P_{\mathrm{R}}$ are the optical power at the photodiode of the probe and reflected wave respectively.

Solving this equation for the reflected power gives 

\begin{equation}\label{eq:Ppr}
    P_\mathrm{R}=16*\frac{P_{\mathrm{RF}}}{R_\mathrm{L}\mathcal{R}_{\mathrm{PD}} {}^2P_{\mathrm{Pr}}}.
\end{equation}

The probe power received by the photodiode is half the sent into the chip, due to the additional 50:50 splitter, so 3.59~dBm. Using Eq.~\eqref{eq:Ppr} gives a $P_{\mathrm{R}}$ of -149.7~dBm at the photodiode. Further taking into account the loss of the 50:50 splitter, the 5~dB insertion loss of the band-pass filter, and the coupling loss, we can calculate that the reflected signal on chip is -125.2~dBm. 

\section*{Discussion and conclusion}

To our knowledge, we present here first BDG signal observed in a silicon nitride waveguide. The probe wavelength that shows the BDG peak matches with our expectations based on theory.

The clear signal measured shows that even when the waveguides are optimized for only one polarization, a BDG can be observed in this platform. An optimized design, which enables better coupling and lower losses for the TM mode will increase the strength of the BDG, and unlock the use of this unique filter in the silicon nitride platform.

\begin{figure}
    \centering
    \includegraphics[width=\linewidth]{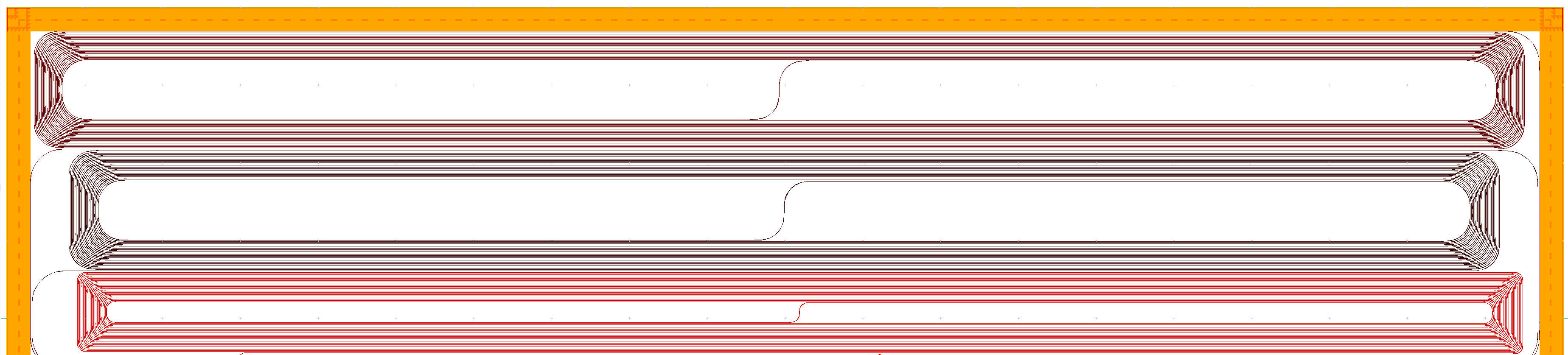}
    \caption{\textbf{The layout of our newly designed sample.} The top two spirals have a bend radius of 300~µm, larger than the standard 100~µm, which can be seen in the bottom spiral. The top waveguides are also connected to a different fiber coupling taper, which has been optimized for the TM mode. This will enhance the BDG response in these devices compared to those used in this work.}
    \label{fig:Spirals}
\end{figure}

To that end we have designed a new sample for our BDG and SBS experiments, part of which can be seen in Fig.~\ref{fig:Spirals}. This sample, currently in fabrication, will contain multiple spirals with different applications. The top two spirals are optimized for further BDG measurements These have different tapers, that are optimized for the TM polarization, and a larger bend radius of 300~µm. This will lower the bend losses for the TM mode to 0.015~dB/cm, which will allow the TM mode to travel along the whole waveguide. These waveguides are 1~m long, which enables us the freedom to tune the BDG length over a large range, and also the ability to add a long delay into a BDG reflected signal. These samples will enable us to test and show the many applications for a BDG in silicon nitride waveguides.

\section*{Author contributions}
D.M., R.B. and J.vd.H developed the concept and proposed the physical system. R.B., J.vd.H. and M.H. developed and performed the numerical simulations. R.B., J.vd.H. and A.M. performed the experiments with input from K.Y., A.v.R. and K.B. R.B. and J.vd.H. wrote the manuscript with input from A.M., A.v.R., K.B. and D.M. D.M. led and supervised the entire project. 

\section*{Acknowledgements}

The authors would like to thank Arjan Meijerink, for his work in designing the new sample.

\section*{Funding Information}

This project is funded by the Nederlandse Organisatie voor Wetenschappelijk Onderzoek (NWO), project numbers 15702 and 740.018.021.

\bibliographystyle{IEEEtran}
\bibliography{library}

\begin{thebibliography}{10}
\providecommand{\url}[1]{#1}
\csname url@samestyle\endcsname
\providecommand{\newblock}{\relax}
\providecommand{\bibinfo}[2]{#2}
\providecommand{\BIBentrySTDinterwordspacing}{\spaceskip=0pt\relax}
\providecommand{\BIBentryALTinterwordstretchfactor}{4}
\providecommand{\BIBentryALTinterwordspacing}{\spaceskip=\fontdimen2\font plus
\BIBentryALTinterwordstretchfactor\fontdimen3\font minus
  \fontdimen4\font\relax}
\providecommand{\BIBforeignlanguage}[2]{{%
\expandafter\ifx\csname l@#1\endcsname\relax
\typeout{** WARNING: IEEEtran.bst: No hyphenation pattern has been}%
\typeout{** loaded for the language `#1'. Using the pattern for}%
\typeout{** the default language instead.}%
\else
\language=\csname l@#1\endcsname
\fi
#2}}
\providecommand{\BIBdecl}{\relax}
\BIBdecl

\bibitem{Aspelmeyer_RevModPhys_2014}
M.~Aspelmeyer, T.~J. Kippenberg, and F.~Marquardt, ``{Cavity optomechanics},''
  \emph{Reviews of Modern Physics}, vol.~86, no.~4, pp. 1391--1452, 2014.

\bibitem{Safavi_Opt_2019}
A.~H. Safavi-Naeini, D.~{Van Thourhout}, R.~Baets, and R.~{Van Laer},
  ``{Controlling phonons and photons at the wavelength-scale: silicon photonics
  meets silicon phononics},'' \emph{Optica}, vol.~6, no.~2, pp. 213--232, 2019.

\bibitem{Eggleton_NatPhot_2019}
B.~J. Eggleton, C.~G. Poulton, P.~T. Rakich, M.~J. Steel, and G.~Bahl,
  ``{Brillouin integrated photonics},'' \emph{Nature Photonics}, vol.~13,
  no.~10, pp. 664--677, 2019.

\bibitem{Li_NatComm_2013}
J.~Li, H.~Lee, and K.~J. Vahala, ``{Microwave synthesizer using an on-chip
  Brillouin oscillator},'' \emph{Nature Communications}, vol.~4, p. 2097, 2013.

\bibitem{Marpaung_Optica_2015}
D.~Marpaung, B.~Morrison, M.~Pagani, R.~Pant, D.-Y. Choi, B.~Luther-Davies,
  S.~J. Madden, and B.~J. Eggleton, ``{Low-power, chip-based stimulated
  Brillouin scattering microwave photonic filter with ultrahigh selectivity},''
  \emph{Optica}, vol.~2, no.~2, p.~76, 2015.

\bibitem{Barzanjeh_NatPhys_2022}
S.~Barzanjeh, A.~Xuereb, S.~Gr{\"{o}}blacher, M.~Paternostro, C.~A. Regal, and
  E.~M. Weig, ``{Optomechanics for quantum technologies},'' \emph{Nature
  Physics}, vol.~18, no.~1, pp. 15--24, 2022.

\bibitem{Dong_OL_2014}
Y.~Dong, T.~Jiang, L.~Teng, H.~Zhang, L.~Chen, X.~Bao, and Z.~Lu, ``{Sub-MHz
  ultrahigh-resolution optical spectrometry based on Brillouin dynamic
  gratings},'' \emph{Optics Letters}, vol.~39, no.~10, p. 2967, 2014.

\bibitem{Kim_JLT_2015}
Y.~H. Kim and K.~Y. Song, ``{Characterization of Nonlinear Temperature
  Dependence of Brillouin Dynamic Grating Spectra in Polarization-Maintaining
  Fibers},'' \emph{Journal of Lightwave Technology}, vol.~33, no.~23, pp.
  4922--4927, 2015.

\bibitem{Li_Opt_2017}
J.~Li, K.~Vahala, and M.-G. Suh, ``{Microresonator Brillouin gyroscope},''
  \emph{Optica}, vol.~4, no.~3, pp. 346--348, 2017.

\bibitem{Li_Optica_2018}
B.-B. Li, J.~B{\'{i}}lek, U.~B. Hoff, L.~S. Madsen, S.~Forstner, V.~Prakash,
  C.~Sch{\"{a}}fermeier, T.~Gehring, W.~P. Bowen, and U.~L. Andersen,
  ``{Quantum enhanced optomechanical magnetometry},'' \emph{Optica}, vol.~5,
  no.~7, p. 850, 2018.

\bibitem{Chin_LPR_2012}
S.~Chin and L.~Th{\'{e}}venaz, ``{Tunable photonic delay lines in optical
  fibers},'' \emph{Laser {\&} Photonics Reviews}, vol.~6, no.~6, pp. 724--738,
  2012.

\bibitem{Merklein_JoO_2018}
M.~Merklein, B.~Stiller, and B.~J. Eggleton, ``{Brillouin-based light storage
  and delay techniques},'' \emph{Journal of Optics}, vol.~20, no.~8, p. 083003,
  2018.

\bibitem{Sancho_OE_2012}
J.~Sancho, N.~Primerov, S.~Chin, Y.~Antman, A.~Zadok, S.~Sales, and
  L.~Th{\'{e}}venaz, ``{Tunable and reconfigurable multi-tap microwave photonic
  filter based on dynamic Brillouin gratings in fibers},'' \emph{Optics
  Express}, vol.~20, no.~6, p. 6157, 2012.

\bibitem{Santagiustina_SciRep_2013}
M.~Santagiustina, S.~Chin, N.~Primerov, L.~Ursini, and L.~Th{\'{e}}venaz,
  ``{All-optical signal processing using dynamic Brillouin gratings},''
  \emph{Scientific Reports}, vol.~3, no.~1, p. 1594, 2013.

\bibitem{Song_OL_2008}
K.~Y. Song, W.~Zou, Z.~He, and K.~Hotate, ``{All-optical dynamic grating
  generation based on Brillouin scattering in polarization-maintaining
  fiber},'' \emph{Optics Letters}, vol.~33, no.~9, p. 926, 2008.

\bibitem{Song_OL_2011}
K.~Y. Song, ``{Operation of Brillouin dynamic grating in single-mode optical
  fibers},'' \emph{Optics Letters}, vol.~36, no.~23, p. 4686, 2011.

\bibitem{Li_OL_2012}
S.~Li, M.-J. Li, and R.~S. Vodhanel, ``{All-optical Brillouin dynamic grating
  generation in few-mode optical fiber},'' \emph{Optics Letters}, vol.~37,
  no.~22, p. 4660, 2012.

\bibitem{Pant_OL_2013}
R.~Pant, E.~Li, C.~G. Poulton, D.-Y. Choi, S.~Madden, B.~Luther-Davies, and
  B.~J. Eggleton, ``{Observation of Brillouin dynamic grating in a photonic
  chip},'' \emph{Optics Letters}, vol.~38, no.~3, p. 305, 2013.

\bibitem{Botter_SciAdv_2022}
R.~Botter, K.~Ye, Y.~Klaver, R.~Suryadharma, O.~Daulay, G.~Liu, J.~van~den
  Hoogen, L.~Kanger, P.~van~der Slot, E.~Klein, M.~Hoekman, C.~Roeloffzen,
  Y.~Liu, and D.~Marpaung, ``{Guided-acoustic stimulated Brillouin scattering
  in silicon nitride photonic circuits},'' \emph{Science advances}, vol.~8,
  no.~40, p. eabq2196, 2022.

\bibitem{Roeloffzen_JSTQE_2018}
C.~G.~H. Roeloffzen, M.~Hoekman, E.~J. Klein, L.~S. Wevers, R.~B. Timens,
  D.~Marchenko, D.~Geskus, R.~Dekker, A.~Alippi, R.~Grootjans, A.~van Rees,
  R.~M. Oldenbeuving, J.~P. Epping, R.~G. Heideman, K.~W{\"{o}}rhoff,
  A.~Leinse, D.~Geuzebroek, E.~Schreuder, P.~W. van Dijk, I.~Visscher,
  C.~Taddei, Y.~Fan, C.~Taballione, Y.~Liu, D.~Marpaung, L.~Zhuang,
  M.~Benelajla, and K.-J. Boller, ``{Low-Loss Si$_3$N$_4$ TriPleX Optical
  Waveguides: Technology and Applications Overview},'' \emph{IEEE Journal of
  Selected Topics in Quantum Electronics}, vol.~24, no.~4, pp. 1--21, 2018.

\bibitem{Daulay_NatComm_2022}
O.~Daulay, G.~Liu, K.~Ye, R.~Botter, Y.~Klaver, Q.~Tan, H.~Yu, M.~Hoekman,
  E.~Klein, C.~Roeloffzen, Y.~Liu, and D.~Marpaung, ``{Ultrahigh dynamic range
  and low noise figure programmable integrated microwave photonic filter},''
  \emph{Nature Communications}, vol.~13, no.~1, p. 7798, 2022.

\bibitem{Liu_APLPhot_2023}
G.~Liu, K.~Ye, O.~Daulay, Q.~Tan, H.~Yu, and D.~Marpaung, ``{Linearized
  integrated microwave photonic circuit for filtering and phase shifting},''
  \emph{APL Photonics}, vol.~8, no.~5, 2023.

\bibitem{Bergman_Sensors_2018}
A.~Bergman and M.~Tur, ``{Brillouin Dynamic Gratings—A Practical Form of
  Brillouin Enhanced Four Wave Mixing in Waveguides: The First Decade and
  Beyond},'' \emph{Sensors}, vol.~18, no.~9, p. 2863, 2018.

\bibitem{Boyd_Book_2008}
R.~W. Boyd, \emph{{Nonlinear Optics}}.\hskip 1em plus 0.5em minus 0.4em\relax
  Elsevier Science Publishing Co Inc, 2008.

\bibitem{Agrawal_Book_2013}
G.~P. Agrawal, \emph{{Nonlinear Fiber Optics}}, 5th~ed.\hskip 1em plus 0.5em
  minus 0.4em\relax Elsevier Science Publishing Co Inc, 2013.

\bibitem{Dong_OE_2010}
Y.~Dong, L.~Chen, and X.~Bao, ``{Characterization of the Brillouin grating
  spectra in a polarization-maintaining fiber},'' \emph{Optics Express},
  vol.~18, no.~18, p. 18960, 2010.

\bibitem{Oldenbeuving_LPL_2013}
R.~M. Oldenbeuving, E.~J. Klein, H.~L. Offerhaus, C.~J. Lee, H.~Song, and K.-J.
  Boller, ``{25 kHz narrow spectral bandwidth of a wavelength tunable diode
  laser with a short waveguide-based external cavity},'' \emph{Laser Physics
  Letters}, vol.~10, no.~1, p. 015804, 2013.

\bibitem{Fan_IEEEPhot_2016}
Y.~Fan, J.~P. Epping, R.~M. Oldenbeuving, C.~G.~H. Roeloffzen, M.~Hoekman,
  R.~Dekker, R.~G. Heideman, P.~J.~M. van~der Slot, and K.-J. Boller,
  ``{Optically Integrated InP–Si$_3$ N$_4$ Hybrid Laser},'' \emph{IEEE
  Photonics Journal}, vol.~8, no.~6, pp. 1--11, 2016.

\bibitem{VanRees_OE_2020}
A.~van Rees, Y.~Fan, D.~Geskus, E.~J. Klein, R.~M. Oldenbeuving, P.~J.~M.
  van~der Slot, and K.-J. Boller, ``{Ring resonator enhanced mode-hop-free
  wavelength tuning of an integrated extended-cavity laser},'' \emph{Optics
  Express}, vol.~28, no.~4, p. 5669, 2020.

\end{thebibliography}

\end{document}